\documentclass[psfig,times]{aa}  
\usepackage{rotating}
\usepackage{psfig}
\begin{document}

\title{
The FORS Deep Field: Field selection, photometric observations
and photometric catalog.
\thanks{Based on observations collected with the VLT$-$UT1 on Cerro Paranal 
(Chile) and the NTT on La Silla (Chile) operated by the European Southern
Observatory in the course of the observing proposals 63.O-0005,
64.O-0149, 64.O-0158, 64.O-0229, 64.P-0150 and 65.O-0048.}
\thanks{Table 4 is only available in electronic form at the CDS via anonymous 
ftp to cdsarc.u-strasbg.fr (130.79.128.5) 
or via http://cdsweb.u-strasbg.fr/cgi-bin/qcat?J/A+A/(vol)/(page)}
}

\author{
Jochen Heidt\inst{1} 
\and
Immo Appenzeller\inst{1}
\and
Armin Gabasch\inst{2}
\and
Klaus J\"ager\inst{3}
\and
Stella Seitz\inst{2}
\and
Ralf Bender\inst{2}
\and
Asmus B\"ohm\inst{3}
\and
Jan Snigula\inst{2}
\and
Klaus J. Fricke\inst{3}
\and
Ulrich Hopp\inst{2}
\and
Martin K\"ummel\inst{4}
\and
Claus M\"ollenhoff\inst{1}
\and
Thomas Szeifert\inst{5}
\and
Bodo Ziegler\inst{3,6}
\and
Niv Drory\inst{2}
\and
D\"orte Mehlert\inst{1}
\and
Alan Moorwood\inst{7}
\and
Harald Nicklas\inst{3}
\and
Stefan Noll\inst{1}
\and
Roberto P. Saglia\inst{2}
\and
Walter Seifert\inst{1}
\and
Otmar Stahl\inst{1}
\and
Eckhard Sutorius\inst{1,8}
\and
Stefan J. Wagner\inst{1}
}
\offprints{J. Heidt,\\
\email{jheidt@lsw.uni-heidelberg.de}}

\institute{Landessternwarte Heidelberg, K\"onigstuhl,
D$-$69117 Heidelberg, Germany
\and
Universit\"atssternwarte M\"unchen, Scheinerstr. 1, D$-$81679 M\"unchen,
Germany 
\and 
Universit\"ats-Sternwarte G\"ottingen, Geismarlandstr. 11, D$-$37083
G\"ottingen, Germany
\and
Max-Planck-Institut f\"ur Astronomie, K\"onigstuhl 17, D$-$69117 Heidelberg,
Germany
\and
European Southern Observatory Santiago, Alonso de Cordova 3107, Santiago
19, Chile
\and
Akademie der Wissenschaften, Theaterstr. 7, 37079 G\"ottingen, Germany
\and
European Southern Observatory, Karl-Schwarzschild-Str. 2, 85748 Garching,
Germany
\and
Royal observatory Edinburgh, Blackford Hill, Edinburgh EH9 3HJ, United Kingdom
}

\date{ Received 11 July 2002 / Accepted 15 October 2002}

\abstract{The FORS Deep Field project is a multi-colour, multi-object 
spectroscopic investigation of a $\sim 7\arcmin \times 7\arcmin$ 
region near the south galactic pole based mostly on
observations carried out with the FORS instruments attached 
to the VLT telescopes. It includes the QSO Q 0103-260 (z = 3.36).
The goal of this study is to improve our understanding of 
the formation and evolution  of galaxies in the young Universe.
In this paper the field selection, the photometric observations, and the data 
reduction are described. The source detection and photometry of objects in the
FORS Deep Field is discussed in detail. A combined B and I selected UBgRIJKs 
photometric catalog of 8753 objects in the FDF is presented and its 
properties are briefly discussed. The formal 50\% completeness limits  
for point sources, derived from the co-added images,
are 25.64, 27.69, 26.86, 26.68, 26.37, 23.60 and 21.57
in U, B, g, R, I, J and Ks (Vega-system), respectively. A comparison of 
the number counts in the FORS Deep Field to those derived in 
other deep field surveys shows very good agreement. 
}

\titlerunning{The FORS Deep Field}
\authorrunning{J. Heidt et al.}

\maketitle

\section{Introduction}

Deep field studies have become one of the most powerful tools to explore 
galaxy evolution over a wide redshift range. One of the main aims of 
this kind of study is to constrain current evolutionary scenarios  
for galaxies, such as the hierarchical structure formation typical of 
Cold Dark Matter universes.

Undoubtedly, the Hubble Deep Field North (HDF-N, Williams et al. 
\cite{hdfn}) and follow-up observations with Keck 
were of particular importance to improve our knowledge of galaxy 
evolution in the redshift range z = 1 - 4
(see e.g. the contributions to the HDF symposium, 1998, ed. Livio et al.). 
The HDF-N is the deepest multi-colour view of the sky made so far, 
with excellent resolution. A disadvantage of the HDF-N
(and its southern counterpart, the Hubble Deep Field
South (HDF-S, Williams et al. \cite{hdfs})) 
is a relatively small field of view ($\sim$ 5.6 sq.arcmin).
Therefore, its statistical results may be affected by the 
large-scale structure (Kajisawa \& Yamada \cite{kajisawa}; 
see also Cohen \cite{cohen}) and by limitations due to small samples. 

\begin{figure*}[t]
\centerline{\hbox{
\psfig{figure=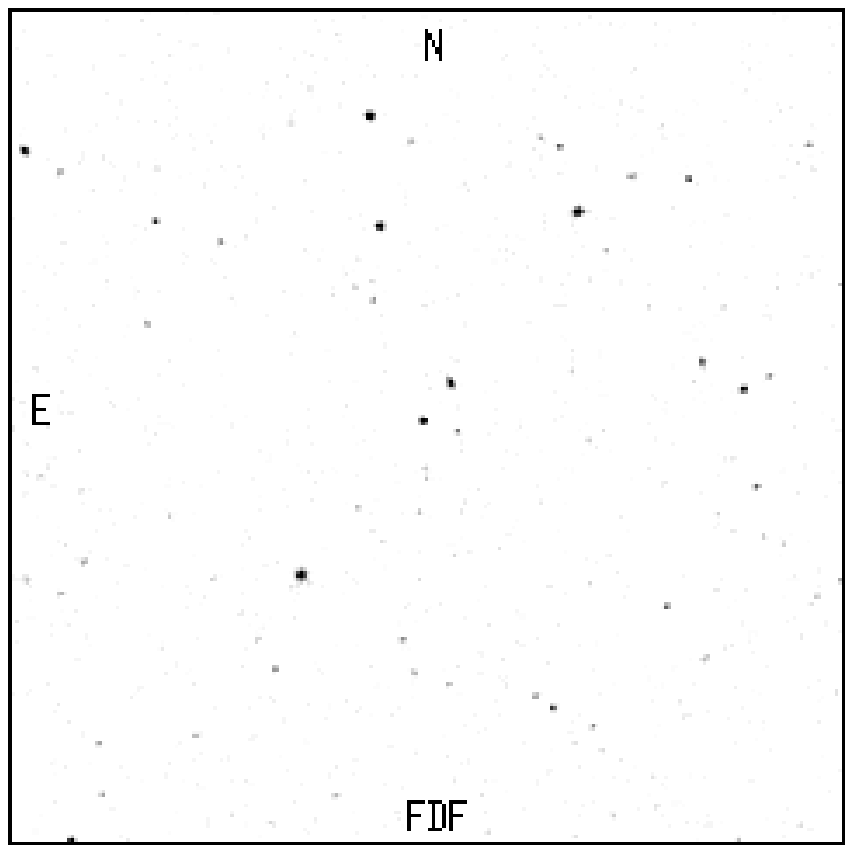,width=5.7cm,height=5.5cm,clip=t}
\hspace*{.2cm}
\psfig{figure=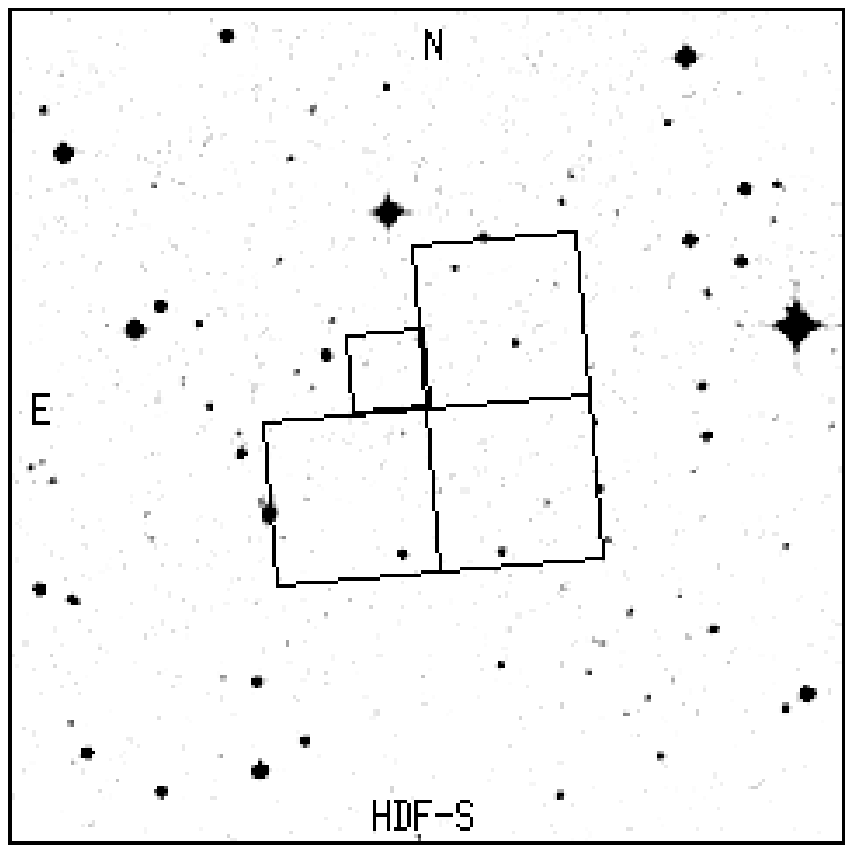,width=5.7cm,height=5.5cm,clip=t}
}}
\caption [] {DSS plots of the FDF and of a field of the same size surrounding
the HDF-S. Also indicated are the field boundaries of the HDF-S.
Note the much lower surface density of bright foreground 
objects and the absence of bright stars in the FDF region.}
\end{figure*}

Following the pioneering work of Tyson (\cite{tyson}) several 
ground-based deep fields with a wide range of scientific drivers, 
sizes, limiting  magnitudes and resolutions have been initiated. 
Examples are the NTT SUSI Deep Field (NTTDF, Arnouts et al. 
\cite{nttdf}), which has a size similar to the HDFs and sub-arcsecond 
resolution, but is a few magnitudes less deep than the HDFs, or 
the William Herschel Deep Field 
(WHTDF, Metcalfe et al. \cite{whtdf} and references therein), which
has a much larger field of view, a depth comparable to the HDFs, but lacks
sub-arcsecond resolution. Other surveys, such as the Calar Alto
Deep Imaging Survey (CADIS, Meisenheimer et al. \cite{cadis}), are 
much shallower, but cover  much larger areas (several 100 sq. arcmin 
in the case of CADIS ) and are specifically designed to search for primeval 
galaxies in the redshift range z = 4.6 - 6.7. 

The aim of the FORS Deep Field (FDF) is to merge some of the strengths 
of the deep field studies cited above. The FDF programme has been carried out 
with the ESO VLT and the FORS instruments 
(Appenzeller et al. \cite{appenzeller98}) at a site, that
offers excellent seeing conditions and allows imaging to almost the depths
of the HDFs. The larger field of view compared to the HDFs
(about 4 times the combined HDFs) alleviates the problem of the 
large-scale structure and results in larger samples of interesting objects.
Moreover, spectroscopic 
follow-up studies with FORS can make full use of the entire field. 
Using the FORS2 MXU-facility, up to $\sim$ 60 spectra of galaxies 
(within 40 slitlets) in the FDF can be taken simultaneously.

In the present paper, the field selection of the FDF, the photometric 
observations and the data reduction are described. The first results have been
described in J\"ager et al. (\cite{jaeger}). A source catalog
(available electronically) based on 
objects detected in the B and I bands and
containing 8753 objects in the FDF is described and its
properties are discussed. This catalog supersedes a preliminary I-band 
selected catalog, which had been discussed by Heidt et al. (\cite{heidt}).
Photometric redshifts obtained from the FDF will be 
discussed by Gabasch et al. (in prep.; see Bender et al. \cite{bender2001}
for preliminary results). Spectroscopic follow-up observations of a subsample
of the FDF galaxies have been started. Up to now, spectra of about 500 
galaxies with redshifts up to z $\sim$ 5 have been analyzed. 
Initial results have been described in Appenzeller et al. 
(\cite{appenzeller02}), 
Mehlert et al. (\cite{mehlert2001}, \cite{mehlert2002}), 
Noll et al. (\cite{noll}) and Ziegler et al. (2002).

\section{Field selection}

A critical aspect for a deep field study is the selection of a suitable
sky area. Since  we intended to obtain a representative 
deep cosmological probe of the Universe, one condition was that the 
galaxy number counts were not disturbed by a galaxy cluster in the field.
To go as deep as possible also requires low galactic extinction
(E(B-V) $<$ 0.02 mag). For the same reason, the field had to be devoid of
strong radio or x-ray sources (potentially indicating the presence of
 galaxy clusters at medium redshifts).
On the other hand, we decided to include a high-redshift (z $>$ 3)
radio-quiet QSO to study the IGM along the line-of-sight to the QSO and the
QSO environment. To facilitate the observations in other wavebands, low HI
column density ($<$ 2 $\times 10 ^{20} {\rm cm}^{-2}$) 
and low FIR cirrus emission was required. Moreover, stars brighter
than 18th mag had to be absent to allow reasonably long exposures,
to avoid saturation of the CCD
and to minimize readout time losses. Because of the latter conditions, 
the HDF-S region was not suitable for our study (see Fig. 1).
Additionally, stars brighter than 5th mag
within $5\degr$ of the field had to be absent to avoid possible reflexes
and stray-light from the telescope structure. Finally, the field had to have
a good observability and, therefore, had to pass close to the 
zenith at the VLT site.

\begin{table}[]
\caption[]{Characteristics of the FORS Deep Field}
\begin{center}
\begin{tabular}{lc}
\hline
 & \\
Field center & $1^{\rm h}6^{\rm m}3\fs6\ \  -25\degr45\arcmin46\arcsec$ 
(2000)\\
mean ${\rm E}({\rm B}-{\rm V})$   & 0.018 \\
H I column density & 1.92 $\times 10^{20} {\rm cm}^{-2}$ \\
Radio sources (NVSS) & none with flux $>$ 2.5 mJy\\
${\rm IRAS Cirrus} (100 \mu{\rm m})$ & 0.035 Jy \\
Bright stars ($<$5 mag)& none within $5\degr$ \\
 & \\
\hline
\end{tabular}
\end{center}
\end{table}

\begin{table*}[]
\caption[]{Observing log of the FDF observations}
\begin{center}
\begin{tabular}{lcll}
\hline
& & & \\
Tel./Inst. & Dates & Filters & Comments\\
& & & \\
\hline
& & & \\
FORS1/UT1 & Aug. 13-17 1999 & g, R & mostly non-phot.\\
FORS1/UT1 & Oct. 6-13 1999 & U, B, g, R, I & during 3 nights\\
FORS1/UT1 & Nov. 3-6 1999  & U, B, R, I & 3 $\times$ 0.5 nights\\
FORS1/UT1 & Dec. 2-6 1999  & U, B, R, I & 4 $\times$ 0.3 nights\\
FORS1/UT1 & July/Aug. 2000 & B, I & 3.5 hours each \\
& & & \\
SofI/NTT & Oct. 25-28 1999 & J, Ks & \\
& & & \\
\hline
\end{tabular}
\end{center}
\end{table*}

Due to these constraints, the south galactic pole region was  
searched for a suitable field. We started by selecting
all the QSOs from the catalog of V\'{e}ron-Cetty
\& V\'{e}ron (7th edition, \cite{veron}) with z $>$ 3 within $10^{\circ}$ 
of the south galactic pole. This resulted in 32 possible field candidates.
Next we did an extensive search in the
literature from radio up to the x-ray regime (FIRST, IRAS maps,
RASS etc.), checked visually the digitized sky survey and used the
photometry provided by the COSMOS scans to select 4 promising field
candidates containing a z $>$ 3 QSO. For these 4 field candidates
short test observations were carried out 
during the commissioning phase of FORS1, which showed that
3 of them were not useful (they either contained
conspicuous galaxy clusters or, in one case, did not provide suitable 
guide stars for the active optics of the VLT).
Finally, a field with the
center coordinates $\alpha_{2000} = 1^{\rm h} 6^{\rm m} 3\fs6,
\delta_{2000} = -25\degr45\arcmin46\arcsec$ containing the QSO Q 0103-260
(z = 3.36, Warren et al. \cite{warren}) was chosen as the FDF. 
The characteristics of this field are summarized in Table 1. 
The Digital Sky Survey (DSS) prints in Fig. 1 provides a
comparison of the FDF and the HDF-S, showing the great advantage of the FDF in
relation to the HDF-S concerning the presence of bright stars.

\section{Observations}

Photometric observations using Bessel UBRI and Gunn g broad band filters 
were carried out with FORS1 at the ESO-VLT UT1 during 5 observing runs in 
visitor mode between August and December 1999. The data were complemented 
with some additional service-mode observations in the Bessel B and I 
filters with the same telescope in July and August 2000. 
Observing conditions were mostly photometric except for the August 1999 
run, which was hampered by the presence of 
clouds and strong winds during some of the nights.
In all cases a 2 $\times$ 2 k TEK CCD in standard
resolution mode ($0\farcs2$/pixel, FOV $6\farcm8 \times  6\farcm8$), 
low gain and 4-port readout was used. 
The Gunn g filter was chosen instead of Bessel V in order to avoid
the 5577 ${\rm \AA}$\ night sky emission line, thus reducing the
background significantly.

From the field-selection images taken with FORS1 it was known 
that twilight flatfields alone are not sufficient for a data reduction
reaching very faint magnitudes. Therefore the images were taken in a 
jittered mode. A 4 $\times$ 4 grid with a spacing of $8\arcsec$
was adopted in order to maximize the use of the scientific images for 
flatfielding purposes on the one hand, and to minimize the loss of 
field-of-view on the other hand. 
The order of the individual observing positions was such 
that images with the largest separation were always taken first. 

Exposure times for the individual frames were set to 1200 sec in U, 
515 sec in B and g, 240 sec in R and 300 sec in I. The seeing limit was 
initially set to $0\farcs5$ for B and I
and $0\farcs8$ for the remaining filters. Unfortunately, it became clear after 
the first observing run that those seeing limits were too strict 
(mainly due to the La Ni\~{n}a phenomenon at that period 
(Sarazin \& Navarrete \cite{sarazin1}, Sarazin \cite{sarazin2}), 
and  could not be met within a reasonable amount of telescope time. 
Therefore the seeing limits were relaxed  to 1$\arcsec$ for 
U and g and $0\farcs8$ for 
the B filter.

Due to the different seeing goals for each filter and varying seeing
conditions during some of the nights, 
images in 3-5 filters were typically taken during each observing run. 
This resulted in somewhat longer
exposure times on the summed images than initially anticipated
(see section 5). Photometric standards from Landolt (\cite{landolt}) 
were taken  at least once during each photometric night.

NIR observations  of the FDF in the J and Ks filter bands 
were acquired using SofI at the ESO NTT 
during 3 photometric nights in October 1999.
Since the field-of-view of SofI with the large field objective 
is $4\farcm94 \times 4\farcm94$ ($0\farcs292$/pixel) only and, thus, 
significantly smaller than the field-of-view offered by FORS1, the observations
were split into 4 subsets to cover the entire FDF.

In order to have as similar observing conditions as possible for all
subsets, the observations in both NIR filters were distributed evenly
over the three nights. Always at least all four subsets were observed
subsequently in one filter for 20 min. Each set of 20 min consisted of
20 exposures of 10 $\times$ 6 sec. The positions of the four subsets were
chosen so as to cover the entire FDF as observed by FORS with a maximal
overlap of the subsets, but to avoid the southernmost 100 pixels of the SofI
camera, which show image degradation (see SofI manual). 
To allow a good sky subtraction, jittered images were taken. 
We used a random walk jitter pattern within a rectangular box of 
$22\arcsec$ border length 
centered on the central position of each subset. Photometric standard
stars from Persson et al. (\cite{persson}) were observed 3 times during 
each night to set the zero point.

In the end, the entire FDF was imaged effectively for 100 min in the two
NIR filters. Due to the overlap of the individual subsets a narrow region was 
observed effectively for 200 min and the central region (including the QSO) 
effectively for 400 min.
An overview of the optical and NIR observing runs and the
filters used is given in Table 2.

\section{Data reduction}

Since we intended to reach with our FDF
observations magnitude limits well below those of earlier ground-based
studies, dedicated data reduction procedures had to be developed. 
On the other hand,
the first spectroscopic follow-up observations of FDF galaxies were to start 
a few months after the last photometric observations of the FDF. In 
order to have candidate galaxies available at that time,
a preliminary reduction of the photometric data taken in visitor mode was made
and an I-band selected catalog with photometric redshifts was created.
The content of this preliminary catalog has been described by 
Heidt et al. (\cite{heidt}), the photometric redshifts for this catalog by 
Bender et al. (\cite{bender2001}). 

In a second step, all data including the photometric data taken in
service mode were reduced as described below. This data set forms the 
basis for the final photometric catalog described in the present paper. 

\subsection{Optical data}
\label{secoptdata}

Because of the time variations of the CCD characteristics and of the
telescope mirror (dust accumulation) each individual run 
was reduced separately. However, in order to have a data set as 
homogeneous as possible, the data reduction strategy was identical for 
all 5 runs.

Firstly, the images were corrected for the bias. Since the observations were
done in 4-port readout mode, each port had to be treated separately.
A masterbias was formed for each port by the scaled median of 
typically 20 bias frames taken during each run, and subtracted from the images
scaling the bias level with the overscan.

Next the images were corrected for the pixel-to-pixel variations and
large-scale sensitivity gradients.
Since the twilight flatfields did not properly correct the large-scale 
gradients, a combination of the twilight flatfields and the science frames 
themselves was used. The twilight flatfields taken in the morning and 
evening generally differed considerably, and the twilight flatfields always
left large-scale gradients on the reduced science frames (probably as a result
of stray-light effects in the telescope and the strong gradient of the sky
background at the beginning and the end of the night). 
Therefore, for each science frame, 
the sequence of flatfields was determined, which minimized the large-scale 
gradient. These sequences were normalized, median filtered and used for 
$1^{\rm st}$ order correction of the pixel-to-pixel variations.
Typically 2-3 flatfields per filter per run had to be created this way,
leaving residuals  of the order of 2-8\% (peak-to-peak) depending on the 
filter. To remove the residuals, the twilight-flatfielded science frames 
were grouped according to similar 2-dim large-scale residuals, normalized
and stacked, using a 1.8 $\sigma$ clipped median. 
Afterwards a correction frame was formed by a 2$-$dim $2^{\rm nd}$ order 
polynomial fit to each median frame. This was done on a 
rectangular grid of 50 $\times$ 50 points, where the level of each grid point
was taken as the median of a box with a width of 40 pixels. 
In this way it was guaranteed that no residuals from stars affected the fit 
and a noise free correction frame was achieved.
Finally, each science frame was corrected for the pixel-to-pixel variations 
by a combination of the corresponding twilight flatfield and noise free
correction frame. The peak-to-peak residuals on the finally reduced
science frames were typically 0.2\% or less.

Cosmic ray events were detected by fitting a two-dimensional
Gaussian to each local maximum in the frame. All signals with a FWHM smaller 
than 1.5 pixels and an amplitude $>$8 times the background noise 
were removed. Then these pixels were replaced by the 
mean value of the surrounding pixels.
This provides a very reliable identification and cleaning of cosmic ray events
(for details see G\"ossl \& Riffeser \cite{goessl}). 

In order to eliminate bad pixels and other affected regions for the
image combination procedure, a bad pixel mask was created for every image. 
The positions of bad pixels on the CCD were determined for each
filter for each run using normalized flatfields. All pixels whose 
flatfield correction exceeded 20\% were flagged. Afterwards,  
each science frame was inspected for other disturbed  regions 
(satellite trails, border effects) and their positions included 
in the corresponding bad pixel masks.

The alignment of the images and the correction for the field distortion 
was done simultaneously. This ensured a minimization of 
smoothing and S/N reduction. As a reference frame, 
an I filter image of the FDF taken under the
best seeing conditions in October 1999 was used. Depending on the filter,
the positions of 15-25 reference stars were measured via a PSF fit on each
frame. A linear coordinate transformation was then calculated to project the
images with respect to the reference image. The transformation included a
rotation, a translation and a global scale variation. Finally, the
correction for the field distortion was applied. Following the ESO
FORS Manual, Version 2.4, we derive the FORS1 distortion corrected
coordinates $(x',y')$ in pixel units as a function of the distorted
coordinates $(x,y)$:
\begin{eqnarray}
\label{eqdist}
x' & = & x-f(r)(x-x_0),\\
y' & = & y-f(r)(y-y_0),
\end{eqnarray}
where $(x_0,y_0)$ are the coordinates of the reference pixel,
$r=\sqrt{(x-x_0)^2+(y-y_0)^2}$ and
\begin{equation}
\label{eqdistfunc}
f(r)=3.602~10^{-4}-1.228~10^{-4}~r+2.091~10^{-9}~r^2.
\end{equation}
The flux interpolation for
non-integer coordinate shifts was calculated from a 16-parameter,
$3^\mathrm{rd}$-order polynomial interpolation using 16 pixel base
points (for details see Riffeser et al. \cite{riffeser}). The same
shifting procedure was applied to the corresponding bad pixel masks,
flagging as 'bad' every pixel affected by bad pixels in the
interpolation. 

The images were then co-added according to the following procedure: First, the 
sky value of each frame was derived via its mode and subtracted.
Then the seeing on each frame was measured using 10 stars,
and the flux of a non-saturated reference star was determined.
Next we assigned a weight to each image relative to the first image in each
filter according to:
\begin{equation}
\label{eqweigth}
weight(n) = {\frac{f(n)}{f(1)}} \times
{\frac{h(1
)\ {\rm FWHM}(1)^2}{h(n)\  {\rm FWHM}(n)^2}}
\end{equation}
where n is the frame to be weighted relative to the $1^{\rm st}$ frame (1), 
f the flux of the reference star, h the sky level and 
FWHM the seeing on the frame. Weights computed according to
Eq. \ref{eqweigth} maximize the signal-to-noise ratio of the combined
image for faint ($f<<h\times{\rm FWHM}^2$) point sources. These are the
overwhelming majority of the objects studied here. 
 Finally, the weighted sum was calculated and normalized to a 1 sec
exposure time. Pixels flagged as bad on the individual images were not
included in the coadding procedure.
Since a different number of dithered frames 
contributed to each pixel in the co-added images, producing a 
position-dependent noise pattern, a combined weight map  
to each frame was constructed. The latter was included into the 
source detection and photometry procedure using SExtractor (see section 6).

The photometric calibration of our co-added frames was done via "reference"
standard stars in the FDF. We first determined the zero points for 
two photometric
nights (Oct. 10/11 and 11/12, 1999) during which the FDF was imaged in all 5
optical filters. The colour correction and extinction coefficients on the ESO 
Web-page were used to derive the zero points for our FORS filter set in
the Vega system. As no calibration images were
available in the $g$-band, transformation from $V$ to $g$ was
performed following J{\o}rgensen (\cite{jorgensen}). 
We then convolved all the FDF images 
from the two photometric  nights to the same seeing as the co-added frames and 
determined the magnitudes of 2 (U) - 10 (I) stars. Based on a curve of growth
for these stars, a fixed aperture with a diameter of $8\arcsec$ was used. 
Using these reference stars, we finally determined the zero points of 
the co-added frames. The difference of the magnitudes between the reference 
stars on the individual frames on the two photometric nights and on the 
co-added frames is 0.01mag or less. We verified our zero points by repeating
the procedure described above using observations from two photometric nights
during our November 1999 run.

\subsection{NIR data}

About $\sim 10 - 20\%$ of the observed NIR frames were found to contain an
electronic pattern caused by the fast motion of the telescope 
near the zenith. These frames were excluded from the analysis.
The remaining data were reduced using standard image processing algorithms
implemented 
within {\sc iraf}\footnote{{\sc iraf} is distributed by the National
  Optical Astronomy Observatories, which are operated by the
  Association of Universities for Research in Astronomy, Inc., under
  cooperative agreement with the National Science Foundation.}.  After
dark-subtraction, for each frame a sky frame was constructed 
typically from the 10 subsequent frames which were scaled to have
the same median counts. These frames were then median-combined using
clipping (to suppress fainter sources and otherwise deviant pixels)
to produce a sky frame. The sky frame was scaled to the median counts of
each image before subtraction to account for variations of sky
brightness on short time-scales. The sky-subtracted images were cleaned
of bad-pixel defects and flat-fielded using dome flats to remove
detector pixel-to-pixel variations. The frames were
then registered to high accuracy, using the brightest $\sim 10$ objects
following the same procedure as described in the previous section,
and finally co-added, after being scaled to airmass zero and an
exposure time of 1 second. 

The additionally observed photometric standard stars were used to
measure the photometric zero point. The typical formal uncertainties in
the zero-points were $0.02$~mag in J and $0.01$~mag in Ks. 

\section{Basic properties of the co-added images}

A summary of the properties of the individual co-added images 
is presented in Table 3. The total integration time for the co-added
images is given as well as the number of frames used, 
the average FWHM measured on 10 stars 
across the field, the area with 80\% weight for each individual image 
and the 50$\%$ completeness limits for a point source
as described in section 6. 

\begin{table*}[]
\caption[]{Overview of the photometric observations.}
\begin{center}
\begin{tabular}{c|ccccc}
\hline
 & & & & & \\
Band & Exposure & Frames & FWHM    & 80\% weight & 50$\%$ compl. limit\\
       & Time [s] & & [\arcsec] &  [$\arcmin^2$] & [mag]     \\
 & & & & & \\
\hline
 & & & & & \\
U       & 44400 & 37  & 0.97 & 40.7 & 25.64\\
B       & 22660 & 44  & 0.60 & 40.5 & 27.69\\
g       & 22145 & 43  & 0.87 & 41.1 & 26.86\\
R       & 26400 & 110 & 0.75 & 40.8 & 26.68\\
I       & 24900 & 83  & 0.53 & 40.9 & 26.37\\
 & & & & & \\
\hline
 & & & & & \\
J       &  4800\footnotemark[2] & 80\footnotemark[2] & 1.20 & 4.2/53.8 & 
23.60/22.85\\
Ks      &  4800\footnotemark[2] & 80\footnotemark[2] & 1.24 & 4.4/53.7 & 
21.57/20.73\\
 & & & & & \\
\hline
\end{tabular}
\end{center}
\begin{minipage}[]{8.5cm}
{$^2$Minimum exposure time and number of frames for each 
subset. Due to the overlap of the subsets for some (small) regions 
of the FDF the total time was  twice or even four times this value.
The 80\% weight and 50\% completeness levels in J and Ks are 
given for the 320 (central field) 
and 80-minutes co-added data, respectively.}
\end{minipage}
\end{table*}

The integration times are in total almost a factor of 2 higher than 
originally planned (except for the U filter). This is due to our strict 
seeing limits during the first observing runs. It compensates, at least in 
part, the loss of resolution/depth of the images
due to the less than optimal seeing. Still, the completeness limits are 
somewhat lower than expected for the integration times since the 
efficiencies of the telescope (reflectivity of the main mirror)
and the CCD were below expected at the time of the observations.
In general, the zero points remained 
relatively constant during the observations carried out in 1999, 
whereas they differed considerably between 
the observations taken in 1999 and 2000.
This resulted in a loss of approx. 0.3~mag (see the ESO-Web page,
Paranal zero points). 

The area with 80\% weight is very similar for all optical bands and 30\%
larger for the NIR bands. The latter is due to the 4 subsets taken during the
NIR observations. The common area with 80\% weight in {\em all} filters
is $39\farcm8^2$.

As an example, the co-added I band image of the FDF is displayed in Fig. 2. 
The common area of the input images for a $6\arcmin \times 6\arcmin$ region 
is shown here. It contains $\sim$ 6100 galaxies. 
In general, the galaxies are distributed
evenly across the field. There is a poor galaxy cluster 
(at z $\sim$ 0.3) in the southwestern corner of the FDF. 
The QSO Q~0103-260 
is south of the center of the frame and is marked with an arrow. 
The brightest object in the field is an elliptical
galaxy with ${\rm m}_{\rm I}$ = 16.5 at z $\sim$ 0.2 
in the southeastern part of the FDF.

\section{Source detection and photometry}

We used SExtractor (Bertin \& Arnouts \cite{bertin})
with the WEIGHT-IMAGE-option and WEIGHT-TYPE $=$ MAP-WEIGHT for
the source detection and extraction on the images. 
The weight-maps described above
were used to account for the spatial dependent noise pattern in the
co-added images, and in particular to pass the local noise
level of the data to the SExtractor program.

\begin{figure*}[t]
\centerline{\hbox{
\psfig{figure=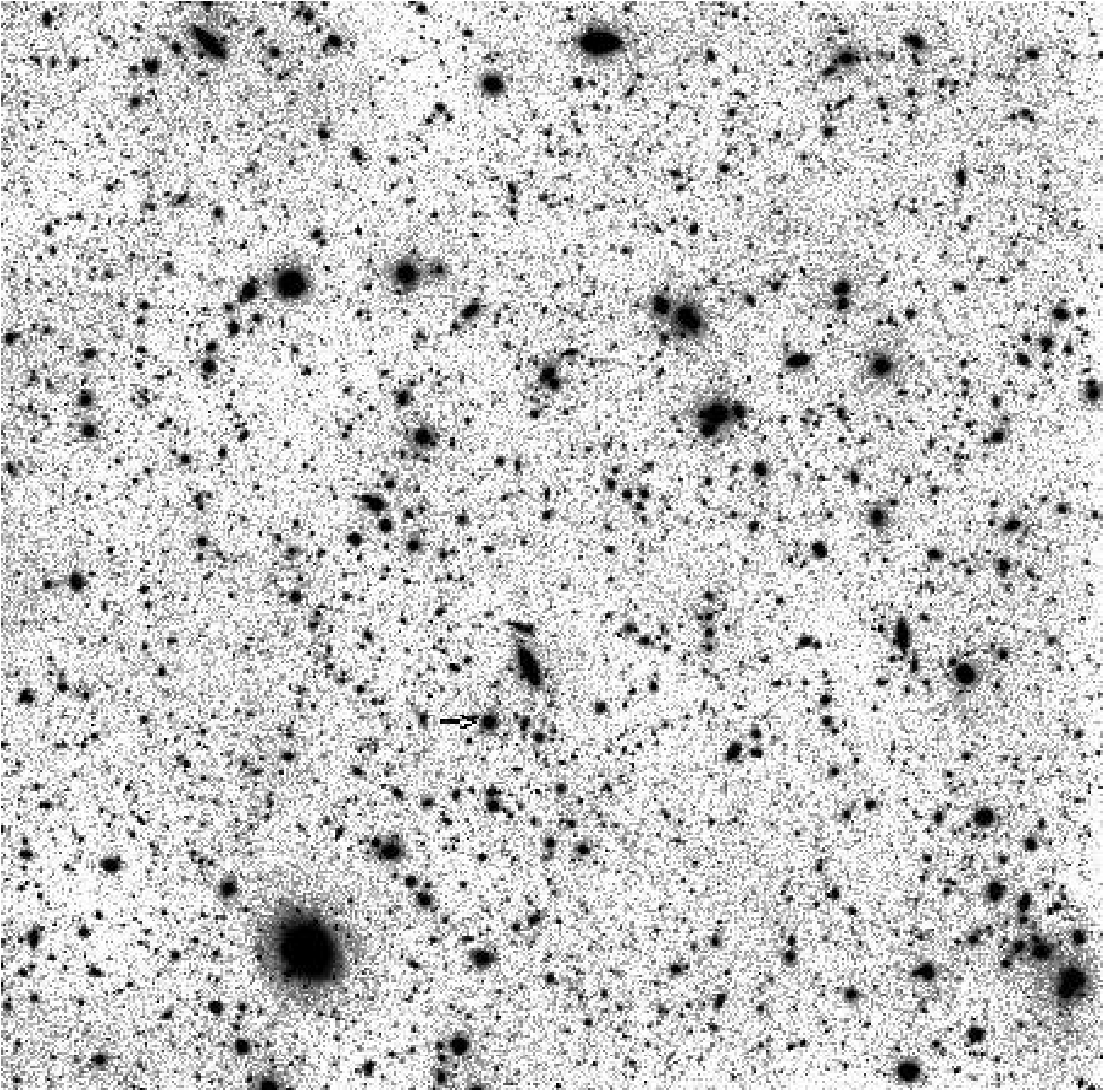,width=18cm,clip=t}
}}
\caption [] {The FDF in I band from FORS observations. The common area
of all input frames for a 
field of view of $6\arcmin \times
6\arcmin$ is shown here. North is up, east to the left. The total integration 
time was 6.9 h, mean FWHM $\sim 0\farcs53$. 
The QSO Q~0103-260 is south of the center of the frame
and marked with an arrow. This area contains $\sim$ 6100 galaxies.
Note the even distribution of galaxies across the
frame, except for the small galaxy concentration in the southwestern corner.
The brightest object in the field is the large elliptical galaxy in the
southeastern part of the FDF at z $\sim$ 0.2 with ${\rm m}_{\rm I} = 16.5$. }
\end{figure*}

To use SExtractor, three parameters have to be set:
i) The detection threshold $t$, which is the minimum 
signal-to-noise ratio of a pixel to be regarded as
a detection, ii) the number $n$ of contiguous pixels exceeding this
threshold, iii) the filtering of the data prior to detection (eg. with a
top-hat or a Gaussian filter). We used a Gaussian filter with a
width $\theta_F$, for the $\theta_F$ values see below.

\begin{figure*}[t]
\centerline{\hbox{
\psfig{figure=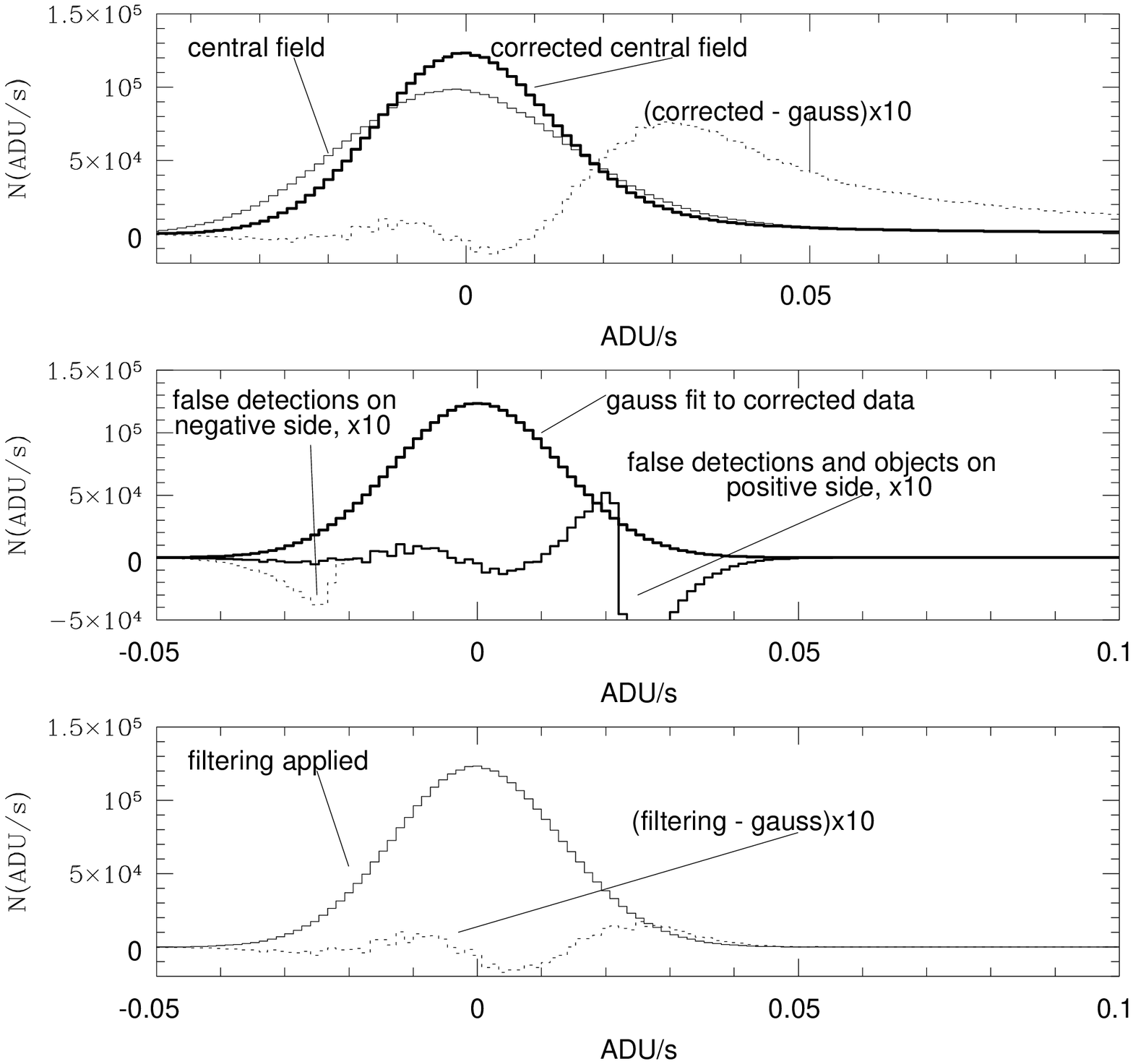,width=13cm,clip=t}
}}
\caption [Optimizing SExtractor extraction parameters] {
Pixel-value histograms (in ADU per second) for the (central field) 
I image at various
analysis stages. {\it Upper panel:} Histogram of the original data (thin
line) and after subtracting the low frequency spatial
variations due to the non-uniform sky background (thick line). Also
included is the difference of the corrected histogram and a Gaussian
(shown as thick line in the {\it middle} panel)
fitted to its negative (ADU/s $<$ 0) wing. This negative wing should
not be affected by
real objects and therefore should represent the true noise in the image.
For clarity the difference has been scaled up by a factor of 10 and
the curve has been labeled accordingly. The real objects show up
as a positive excess of the pixel values in the corrected data
distribution and in the difference function at positive ADU/s.
{\it Middle panel:} The thick line shows the Gaussian derived by
fitting the negative wing of the
corrected data distribution  as described above. Its
difference to the pixel-value distribution derived for those pixels
where SExtractor (with optimal parameters but without filtering)
finds no objects (or object contributions)  is shown as a solid line. The
corresponding difference distribution of the inverted image is shown
dotted for the negative ADU/s only. 
The negative excess shows the false detections due to the
correlated error. 
The difference curves are again scaled up by a factor of 10.
 {\it Lower panel:} 
The thin line shows the histogram of the pixel values of pixels not 
belonging to objects when
SExtractor is run after filtering the corrected data
with a (2 pixel FWHM) Gaussian. The dotted line shows the difference
between this histogram and the Gaussian fit shown in the middle
panel. The number of significant false
detections has now dropped to nearly zero.
}
\label{figdetections}
\end{figure*}

We varied these parameters to maximize the number of source
detections, while minimizing false detections. The following procedure, 
described here for the I-band data, was used for all filters.
We first considered only those pixels in the
field where the exposure time equaled the total exposure time (the
weight-map took care of the correct scaling of RMS for the full field
later on) and called this part of data the `central field'. 

If there were no objects in the field and if the data reduction
resulted in a perfectly flat sky we would expect the histogram of the
pixel-values to be a Gaussian, with a width reflecting the
photon-noise and the correlated noise of the data reduction and
coaddition procedure. The actual histogram of pixel-values of the
central-field is shown in Fig. \ref{figdetections} (upper panel, thin
line). Even ignoring the wings, the histogram is asymmetric around its
center at zero. This stems from the non-uniformities of the sky background,
that amount to about 1\% (see Sect. \ref{secoptdata}).  Therefore, we
determined the sky-curvature on large scales and subtracted
a 2-dimensional fit to this surface from
the original data. The corrected histogram of
pixel-values (Fig. \ref{figdetections}, upper panel, thick curve) is
now symmetric around its center at zero and the left-hand part is well
described by a Gaussian (with a width of $0.01295$ ADU/s). The right
hand part shows an excess above $\approx 0.015$ ADU/s, which is due to
the objects in the field (see difference curve in
Fig. \ref{figdetections}, scaled up by a factor 10).
We have checked that it does not make any
difference for the detection and the photometry of reliable objects
whether the procedure is applied to the original or to the
corrected data: for each object, the difference between
the magnitude estimates of these two cases is smaller than the
assigned magnitude RMS-error. This implies that we can carry out the
adjustment of optimum SExtractor parameters in the
corrected version of the data.

To optimize the pre-detection filtering procedure we made the
following numerical experiment.  We generated a ''negative version''
of an image by multiplying it by $-1$ and a ''randomized version''
by randomly assigning measured pixel values to new positions (the
weights of the weight-map are re-localized the same way).  With no
filtering ($\theta_F=0$) and using t = 1.7 and n = 3 SExtractor finds
about 9000 objects in the original image, 5600 in the negative one and
1100 in the randomized one.  The fact that many more objects are
detected in the negative image than in the randomized one indicates
that correlated noise is present in both the negative and the positive
images. Therefore filtering must be used to specifically suppress 
the small-scale noise. It is possible that large-scale noise is still present,
but there is no way to remove such a component.
  By varying the width $\theta_F$ of a Gaussian filter
we found that $\theta_F = 2$ is an
optimal choice. With $n=3$ and $t=1.7$ the number of objects
detected on the negative image dropped to the expected random number,
nearly zero. Of course, once $\theta_F$ is fixed, one is still left
with the freedom of trading $n$ for $t$ by increasing the number of
pixels above the threshold and decreasing the
threshold value at the same time. 
We decided to keep $n$ small, in order to obtain an
unbiased detection of faint point sources. This choice allows us to
exploit the excellent seeing of the I-band data, where the FWHM is only
$2.5$ pixels.

Now we illustrate our procedure more quantitatively: we ran SExtractor
(for each choice of $\theta_F$, $n$ and $t$) on the positive, the
negative and the randomized images.  We registered all pixels which
were covered by objects, removed them from the pixel-value statistics
and normalized the corresponding pixel-value histogram to the total
number of pixels in the central field, and we call that the
`background-histogram'.  We expect that for good source extraction
parameters, the background histograms will look like a Gaussian, more
precisely like that Gaussian derived by fitting the negative wing of
the corrected data distribution, which we call the
`optimum-background-histogram' below.  The difference (magnified by a
factor of 10) to that optimum background histogram
is shown in the middle panel of
Fig. \ref{figdetections} for $n=3$, $t = 1.7$, $\theta_F=0$
 for detection on the positive (solid) and
negative (dotted, for negative ADU/s only) image. The negative excess
of these histograms below zero are false detections due to correlated
noise.  Increasing $\theta_F$ these false detections drop dramatically
when $\theta_F=2$ pixels is reached. Then, $n=3$ and $t=1.7$ were fixed
by requiring no false detections on the negative image, i.e. no
detections due to correlated noise. We finally run SExtractor with
this set of parameters on the positive image, obtain the background
histogram and show the difference to the optimum background histogram
in the lower panel of Fig. \ref{figdetections} (dotted histogram,
magnified by a factor of 10). The difference is indeed very small.

Using the above parameters ($\theta_F=2$ with a Gaussian
convolution, $n=3$ and $t = 1.7$), obtained from the optimum
pre-detection filtering and the  requirement of
no-detection on the negative image, we find that the extended wing in
the ADU-histogram due to the presence of objects disappears and that
the histogram becomes symmetrical and Gaussian (see
Fig. \ref{figdetections}, bottom panel).  This demonstrates that with
this choice of parameters we are optimally extracting all objects
above the noise level, without getting significant false
detections. The adopted parameters give  a (total) photometric
accuracy better than 5$\sigma$.

The optimum parameters were finally used to run SExtractor on the
(positive and negative) images of the total FDF.  We found about 6900
objects on the positive and less than a handful of objects on the
negative side of the entire I image.  All these spurious detections
occurred near discontinuities of the S/N level outside the central
field and were caused by the non perfectly flat sky, which makes some
of the discontinuities more pronounced than they should be according
to the photon-noise and the corresponding weight-map.

The same analysis described for the I-band image was carried out for
the other filters. We emphasize here that our extraction
procedure was optimized to maximize the number of  real detections for 
a reliable photometry and hence reliable photometric 
redshifts rather than to study
galaxy number counts at the faintest limits.
For the optical bands, we used the same extraction
parameters. For the NIR-data we opted for $\theta_F=3$ pixels to match
the pixel size of the original NIR-data, which is roughly 1.5 the pixel
size of FORS, and $t=2.0$ and $n=5$ for the J band, and  $t=1.9$ and $n=5$
for the Ks band, to take into account the poorer
seeing and the different noise level. To illustrate the reliability of
our detection procedure we display a detection file returned from
SExtractor for a $1\arcmin \times 1\arcmin$ region of the northern
part of the FDF in Fig. 4.

The photometric errors presented in the 
final catalog are those derived by the
SExtractor routine. To make sure that the error calculation was
not influenced by correlated noise in the sky background, the results
of the SExtractor were verified with aperture photometry with
different apertures in areas not covered by objects and by
estimating the expected photometric errors from the background
variations. In general we found good agreement with the SExtractor
derived errors. In particular the SExtractor errors were found
to be quite accurate for point sources and for small objects.
Only in the case of large extended objects may non-stochastic background
variations have resulted in an underestimate of the photometric errors.
But the few objects possibly affected are normally bright
and have small errors, which should still be correct within the numbers
given in the catalog.

Finally, we calculated the 50\% completeness levels in each
filter band using our extraction parameters and the formula given in
Snigula et al. (\cite{snigula}). This approach estimates the
completeness limit by calculating the brightness at which the area of
pixels brighter than the applied flux limit falls below the size
threshold of the detection algorithm (for a given FWHM of a point
source).  To allow a comparison with other deep fields, the data were
corrected for galactic extinction as described in section 7.  The
results are summarized in Table 3.

\begin{figure}[]
\psfig{figure=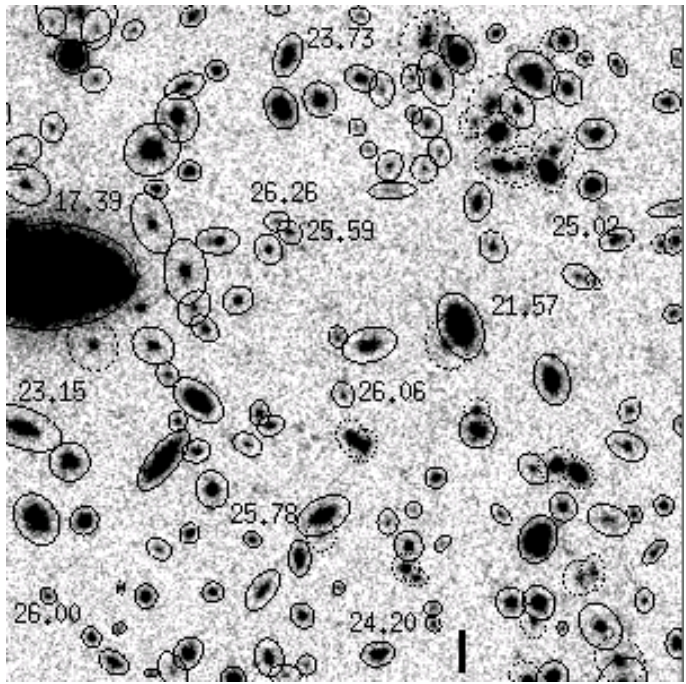,width=9cm,height=9cm,clip=t}
\caption [] {Detection file returned from SExtractor for a
$1\arcmin \times 1\arcmin$ region of the northern part of the FDF.
It illustrates the reliability of our detection and
photometry procedure. The I-band image shown here contains $\sim$ 160 objects.
For some objects the integrated magnitudes are displayed. The detection
file shows the elliptical aperture limits used to derive mag$\_$auto. Dashed
ellipses denote blended objects. 
}
\end{figure}

\section{Photometric catalog}

\subsection{Compilation of the photometric catalog}

To create the final photometric catalog we merged the individual
catalogs of the objects detected in the co-added B filter image and in
the co-added I filter image. We decided to use these two catalogs as a basis, 
since the images in these two filters correspond to the best
seeing conditions and since most types of objects are expected to be
detected in at least one of these two bands.

The merging of the I and B catalogs was carried out as follows: We
first matched the positions of the detected objects and their
corresponding images in the two filters. 
This was done by visual inspection of the entries of
the objects on both frames. This procedure gave us a clear view of
the success of our automatic detection procedure and allowed us to
reject obviously false identifications. In order to avoid
mis-matches in the final catalog, each entry in the B catalog was
first assigned a corresponding entry in the I catalog and vice
versa. A cross-match of the B versus I and I versus B entries allowed
us to identify false matches, which were checked again until a perfect
cross-match was derived.

The initial catalogs in B and I contained 7206 and 6900 entries, respectively.
After the visual cross-matching, we deleted 15 objects from the B catalog
and 8 objects from the I catalog. These were mostly objects close to the edges
of the field.
In a few cases, 2 objects separated by a few pixels (e.g.
a merging pair of galaxies) were detected in the B band, 
whereas in the I band only
one object in between the two B band objects was found (essentially 
at the center of the common envelope of both galaxies). 
In such cases the entry in the I 
band was deleted. This left us with 7191 entries in the B catalog and 6892 
entries in the I catalog. Now we merged both catalogs to form the 
final photometric catalog. This catalog contains 8753 objects. 5327 out of the
8753 objects were detected in both filters (61\%), whereas 1864 (21\%) were 
detected in B only and 1562 (18\%) were detected in I only. We emphasize here 
that a non-detection does not necessarily mean that 
the object is not present on
the frame, it rather means that the object was not detected by SExtractor
with the parameters set here.

Since SExtractor may use a different number of pixels to derive the total
magnitudes in B and I, the colours of very extended objects 
computed from the total magnitudes are not 
reliable.  Therefore the catalog also contains aperture 
magnitudes in UBgRIJKs. An aperture of $2\arcsec$ was chosen 
in order to minimize the errors due to blending and since 
the faint objects usually have diameters of $\leq 2\arcsec$.
The aperture magnitudes were derived by first 
convolving all frames to the same seeing ($1\arcsec$ FWHM) and then 
performing aperture photometry on the positions of the objects detected 
in B and I in the convolved frames. For objects detected in B only, we used
the aperture photometry based on the positions in the B catalog, whereas
the aperture photometry based on the positions in the I catalog were used 
for the remaining objects (detection on both frames or I-only detections).
Thus for many objects, which were initially not detected in either filter,
useful photometric data could be given. 

Finally, the galactic absorption towards the FORS Deep Field was estimated. 
We used the formulae 2 and 3 in Cardelli et al. (\cite{cardelli}) 
and adopted ${\rm E}({\rm B}-{\rm V})$ = 0.018 (Burstein \& Heiles 
\cite{burstein}) and 
$A_V = 3.1 \times {\rm E}({\rm B}-{\rm V})$ to calculate the
extinction correction for each filter. The central
wavelengths for each filter were taken from the ESO Web-page.
We derived $A_U/A_V$ = 1.555, $A_B/A_V$ = 1.365, $A_g/A_V$ = 1.105, $A_R/A_V$ 
= 0.790, $A_I/A_V$ = 0.631, $A_J/A_V$ = 0.283 and $A_{Ks}/A_V$ = 0.117 
resulting in $A_U$ = 0.087 mag, $A_B$ = 0.076 mag, $A_g$ = 0.062 mag, $A_R$ 
= 0.041 mag, $A_I$ = 0.035 mag, $A_J$ = 0.016 mag and $A_{Ks}$ = 0.007 mag, 
respectively. The values for the extinction agree to $\leq$ 0.01 mag
with those listed in the NED.
The photometric catalog described below is not corrected for
galactic extinction. However, the completeness limits as well as the 
number counts shown in section 8 were derived with a galactic extinction
correction.

\subsection{Contents of the photometric catalog}

The full catalog containing 8753 objects is 
available in electronic form at CDS via anonymous ftp to cdsarc.u-strasbg.fr
(130.79.128.5) or via 
http://cdsweb.u-strasbg.fr/cgi-bin/qcat?J/A+A/(vol)/(page).
As an illustration of its content we list in Table 4 
the entries 2630 $-$ 2639.

For each object we report the 
following parameters:

\begin{sidewaystable*}
\centering
\center{\caption[]{Excerpt from the FDF object catalog. The 
entries with the IDs 2630 $-$ 2639 are displayed as examples.}}
\begin{tabular}{rcccrcccccccccccc}
\hline
& & & & & & & & & & & & & & & & \\
ID & RA (2000) &  Dec (2000) & 
${\rm m}_{BT}$  & $\sigma_{\rm BT}$ &
${\rm m}_{IT}$  & $\sigma_{\rm IT}$ &
${\rm m}_{U}$ [$2\arcsec$] & $\sigma_{\rm U}$ &
${\rm m}_{B}$ [$2\arcsec$] & $\sigma_{\rm B}$ &
${\rm m}_{g}$ [$2\arcsec$] & $\sigma_{\rm g}$ &
${\rm m}_{R}$ [$2\arcsec$]& $\sigma_{\rm R}$  &
${\rm m}_{I}$ [$2\arcsec$] & $\sigma_{\rm I}$ \\

& & & & & & & & & & & & & & & & \\
\hline
& & & & & & & & & & & & & & & & \\
 2630 & 1 5 57.28 & -25 48 02.3 & 27.75 & 0.19 & 25.30 & 0.10 & 26.99 & 0.27 & 27.61 & 0.05 & 27.72 & 0.10 & 26.10 & 0.03 & 25.34 & 0.02 \\
 2631 & 1 5 57.29 & -25 45 00.1 &       &      & 24.42 & 0.03 &       &      &       &      & 30.73 & 1.65 & 26.57 & 0.04 & 24.49 & 0.01 \\
 2632 & 1 5 57.29 & -25 48 46.9 & 26.13 & 0.05 & 24.98 & 0.07 & 25.96 & 0.10 & 26.20 & 0.01 & 25.92 & 0.02 & 25.42 & 0.02 & 25.05 & 0.02 \\
 2633 & 1 5 57.30 & -25 44 56.6 & 24.47 & 0.01 & 22.75 & 0.01 & 24.60 & 0.03 & 24.60 & 0.01 & 23.74 & 0.01 & 23.26 & 0.01 & 22.87 & 0.01 \\
 2634 & 1 5 57.30 & -25 48 14.2 & 27.69 & 0.16 &       &      &       &      & 27.77 & 0.06 & 28.23 & 0.17 & 26.84 & 0.06 & 26.78 & 0.09 \\
 2635 & 1 5 57.31 & -25 43 52.3 &       &      & 25.02 & 0.09 & 26.22 & 0.13 & 26.42 & 0.02 & 26.11 & 0.02 & 25.66 & 0.02 & 25.33 & 0.02 \\
 2636 & 1 5 57.31 & -25 44 02.2 & 24.85 & 0.04 & 23.43 & 0.04 & 25.53 & 0.07 & 25.53 & 0.01 & 25.12 & 0.01 & 24.56 & 0.01 & 24.12 & 0.01 \\
 2637 & 1 5 57.31 & -25 44 15.2 & 26.60 & 0.09 & 26.19 & 0.17 & 26.76 & 0.22 & 26.83 & 0.02 & 26.72 & 0.04 & 26.46 & 0.04 & 26.16 & 0.05 \\
 2638 & 1 5 57.31 & -25 46 23.5 & 27.36 & 0.16 & 25.65 & 0.09 & 27.58 & 0.46 & 27.43 & 0.04 & 27.45 & 0.08 & 26.72 & 0.05 & 25.67 & 0.03 \\
 2639 & 1 5 57.31 & -25 47 51.1 & 26.17 & 0.08 & 25.11 & 0.10 & 26.42 & 0.16 & 26.85 & 0.02 & 26.74 & 0.04 & 26.22 & 0.03 & 25.60 & 0.03 \\
& & & & & & & & & & & & & & & & \\
\hline
\end{tabular}

\vspace*{.5cm}

\begin{tabular}{cccccccccccccccc}
\hline
& & & & & & & & & & & & & \\
ID & 
${\rm m}_{J}$ [$2\arcsec$] & $\sigma_{\rm J}$ &
${\rm m}_{Ks}$ [$2\arcsec$] & $\sigma_{\rm Ks}$ &
FWHM [\arcsec] & Elong & PA [\degr] & Cstar & Flag1 
& Flag2 & Flag3 & weight\_B & weight\_I\\
& & & & & & & & & & & & & \\
\hline
& & & & & & & & & & & & & \\
 2630 &       &      & 21.97 &  0.20 &  0.74 & 1.17 &  17.9 & 0.40 & 0 &       &       & 1.000 & 1.000 \\
 2631 & 21.36 & 0.01 & 20.35 &  0.03 &  0.52 & 1.02 & 111.7 & 0.98 & 0 & Ionly & L star&       & 1.000 \\
 2632 & 26.58 & 2.38 & 22.37 &  0.29 &  0.78 & 1.12 &  82.1 & 0.26 & 0 &       &       & 1.000 & 1.000 \\
 2633 & 22.09 & 0.03 & 20.91 &  0.06 &  0.53 & 1.04 &  36.2 & 0.98 & 0 &       &  QSO  & 1.000 & 1.000 \\
 2634 &       &      &       &       &  1.01 & 1.25 &  00.6 & 0.61 & 0 & Bonly &       & 1.000 &       \\
 2635 & 23.70 & 0.18 &       &       &  1.13 & 1.19 & 129.3 & 0.00 & 3 & Ionly &       &       & 0.984 \\
 2636 & 22.71 & 0.07 & 20.75 &  0.07 &  0.73 & 1.34 &  90.2 & 0.09 & 3 &       &       & 0.984 & 1.000 \\
 2637 &       &      &       &       &  1.07 & 1.87 &  76.9 & 0.40 & 0 &       &       & 1.000 & 1.000 \\
 2638 &       &      &       &       &  0.80 & 1.49 &  19.1 & 0.43 & 0 &       &       & 1.000 & 1.000 \\
 2639 & 24.02 & 0.23 & 22.96 &  0.50 &  1.34 & 1.16 &  21.6 & 0.01 & 2 &       &       & 1.000 & 1.000 \\
& & & & & & & & & & & & & \\
\hline
\end{tabular}
\end{sidewaystable*}

ID: The identification number. The objects have been sorted first by  
right ascension (2000), followed by declination (2000). 
The identification numbers provide a cross-reference to the 
spectroscopic and  other observations of the FDF(e.g. Noll et al., in prep).  

RA, Dec: The positions of the objects in the FDF for J2000.0. 
Their accuracy has been examined by comparing the positions of 31 
well-isolated, evenly distributed objects on the I frame of the
FDF, to those listed in the  USNO catalog (Monet \cite{monet}). 
The mean difference in right ascension is $0\farcs21\pm0\farcs38$ and 
the mean difference  in declination is $0\farcs14\pm0\farcs40$. 
Given a typical accuracy of $0\farcs25$ for objects in the USNO catalog 
our positions have an accuracy of $\sim 0\farcs5$.

${\rm m}_{BT}$, $\sigma_{\rm BT}$, ${\rm m}_{IT}$, $\sigma_{\rm IT}$:
The total magnitudes (Vega-system) 
and associated mean errors of the detected sources in the 
B and I band images, respectively, as measured using the SExtractor 
routine mag$\_$auto on the co-added and unconvolved frames. 
Mag$\_$auto is an automatic 
aperture routine based on Kron's (\cite{kron}) "first moment" algorithm, which
determines the sum of counts in an elliptical aperture. The semimajor axis
of this aperture is defined by 2.5 times the first moments of the 
flux distribution within an ellipse roughly twice the isophotal 
radius, within a minimum semimajor axis of 3.5 pixels.
This routine is intended to give the most precise estimate 
of "total magnitudes", at least for galaxies, and takes into account the  
blending of nearby objects.  

${\rm m}_{\rm UBgRIJKs [2\arcsec]}, \sigma_{\rm UBgRIJKs}$: 
UBgRIJKs magnitudes (Vega-System) and associated errors within an aperture
of $2\arcsec$. 
They (and their errors) 
were measured on the co-added and convolved frames
using SExtractor. The positions listed in the catalog were used for this
procedure.
An aperture of $2\arcsec$ was chosen in order to minimize the errors due to
blending. Moreover, the faint objects in the FDF usually have 
diameters of $\leq 2\arcsec$. Choosing a larger aperture would result
in larger photometric errors due to the sky background.
For extended objects, the mean errors of the aperture magnitudes are
generally smaller than for the total magnitudes, as the aperture photometry
selected the regions of high surface brightness.
The magnitudes were not corrected for blending. Blended
objects can be identified from the column Flag1 (see below).

The next four columns (FWHM, elongation, position  angle, star-galaxy
classification parameter) describe the morphology 
of the objects. Since the FWHM, elongation and position angle may have
high errors and are sometimes unreliable for faint objects, 
this information is provided for objects brighter than our 
50$\%$ completeness limit  (27.69 in B, 26.37 in I) only. 
Moreover, we do not list
these values for objects where SExtractor derived a FWHM $<$ 0.4
(FWHM is $0\farcs53$ in co-added I band frame and $0\farcs6$ in 
co-added B band frame). 
The information should also be treated with caution for brighter objects  
having a star-galaxy classification parameter $>$ 0.9. 

FWHM: Full width at half maximum of the objects in arcsec as determined
by SExtractor by a Gaussian fit to the core. 

Elong: Elongation of the images. The elongation is defined as $A/B$,
where A and B are given by the $2^\mathrm{nd}$ order moment of the light
distribution along the major and minor axis, respectively.

PA: The position angle of the major axis, measured from North to East, with 
N-S = 0.

Cstar: Star-galaxy classification parameter returned by SExtractor based on
the morphology of the objects on the image. A classification near 1.0
describes point like sources whereas a classification close to 0.0
describes extended sources.

Flag1: Flags returned by SExtractor with the following notation:

\noindent 1: Object has neighbours bright and close enough 
to bias significantly mag$\_$auto;
2: The object was originally blended with another one;
3: Sum of 1 + 2;
4: At least one pixel of the object is saturated (or very close to saturation);
7: Sum of 1 + 2 + 4;
8: The object is truncated (e.g. too close to the image boundary);
16: Object aperture data are incomplete or corrupted;
17: Sum of 1 + 16;
18: Sum of 2 + 16;
19: Sum of 1 + 2 + 16;
24: Sum of 8 + 16.

Flag2: Here we report if an object was detected on the B frame 
only ("Bonly"), on the I frame only ("Ionly"). If there is no entry, the
object is detected by SExtractor on both frames.

Flag3: A preliminary classification of 35 point-like objects (QSOs, stars) 
from our spectroscopic survey (Noll et al., in prep.). 

weight\_B, weight\_I: Averaged weights of all pixels used to determine
${\rm m}_{BT}$ and ${\rm m}_{IT}$, respectively. They were derived from the
combined weight maps which are described in section 4. 
A weight of 1 means that all
pixels used to derive the magnitude are fully exposed and not affected by bad 
areas. Most of the detections with low weights are close to the edges of 
the FDF where the total integration times are lower.

\section{Galaxy number counts}

The number counts serve as a quick check of the approximate
photometric calibration and for the depth of the data.  We did
not put much effort in star-galaxy separation at the
faint end, where the galaxies dominate the counts anyway.  At the
bright end, where SExtractor is able to disentangle a stellar and a
galaxy profile, we derived limits 
by investigating the class-FWHM diagram for the objects. In the following
figures, the counts for all objects are shown as dashed histograms, while
for the solid line histograms obvious stellar objects have been omitted.
The magnitudes are given in the Vega-system.
The number counts are given only for the area with maximum
integration-times (weight-map $ \approx 1$) for the optical data and
for ${\rm weight-map} \ga 0.25$ for the NIR-data 
(i.e. we exclude the edges of the
fields). They are not corrected for incompleteness. 
Also indicated is the $50$\% completeness limit
for the detection of point sources.
For each filter we also included for comparison 
number-magnitude-relations obtained in earlier 
observations which are compiled  
and transformed to standard filter systems in Metcalfe
et al. (\cite{whtdf}) for the optical filters.
In all cases we plot raw number counts only, i.e. we do not correct
for incompleteness at the faint end.

In the U-band the FDF is 50$\%$ complete to $U = 25.64$ mag for a
point source. The slope agrees with earlier measurements (roughly
$0.4-0.5$) for $U<23$ and it becomes shallower ($0.35$ at $U=23-25$),
in agreement with the slopes of the HDF-S, WHDF and Hogg et al. 1997
(see Metcalfe et al. (\cite{whtdf})). In Fig. 5 we have transformed
the HDF number counts as proposed by Metcalfe et al. using
$F_{3oo,Vega}=U-0.4$ and Table 5 in their paper. We further assume
$U_{WHDF}\approx U$ to include  the WHDF U-band-raw counts
(Table 4 of  Metcalfe et al. 
\cite{whtdf}) --in fact the central wavelengths and the
transmission curves of the U filters used for the FDF and WHDF
observations are similar. The values of Hogg et al. (\cite{hogg}) have been
obtained from their Fig. 3 and been transformed as proposed by Metcalfe,
$U\approx U_{Hogg}+0.08$.  The HDFN/S and WHDF number counts are not corrected
for reddening (Metcalfe, private comm., $E(B-V)_{WHDF}\approx 0.02$
which is similar to the FDF and thus would shift the number counts by
$\approx -0.1$). 

\begin{figure}
\centerline{\hbox{
\psfig{figure=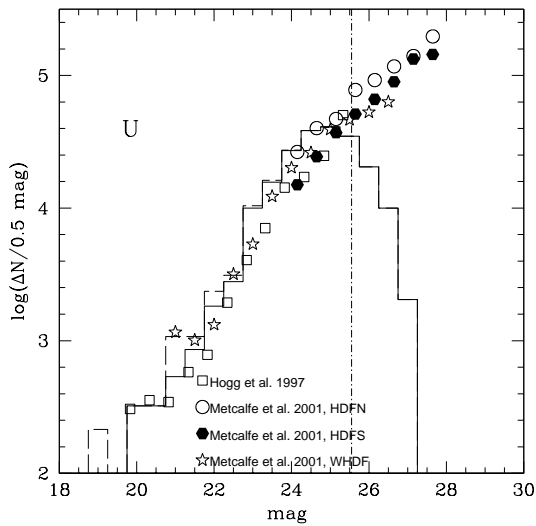,width=7.5cm,angle=0}
}}
\caption [U-band] {Galaxy number counts of the FDF in the U band
(not corrected for incompleteness) as compared
to other deep surveys. The vertical 
dash-dotted line indicates the 50\% completeness limits.}
\end{figure}
Our B-band number counts (Fig. 6) are 50$\%$-complete at $27.69$ mag.
Within the field-to-field variations they agree well with the HDFS/N 
(we follow Metcalfe et al. (\cite{whtdf}) and use the transformation
 $F_{450,Vega} \approx B -0.1$) and the raw-counts in  the NTT deep field 
(Arnouts et al., \cite{nttdf}). We also included the raw counts in the
Herschel deep field, assuming $B_{FDF}\approx B_{WHDF}$. 
\begin{figure}
\centerline{\hbox{
\psfig{figure=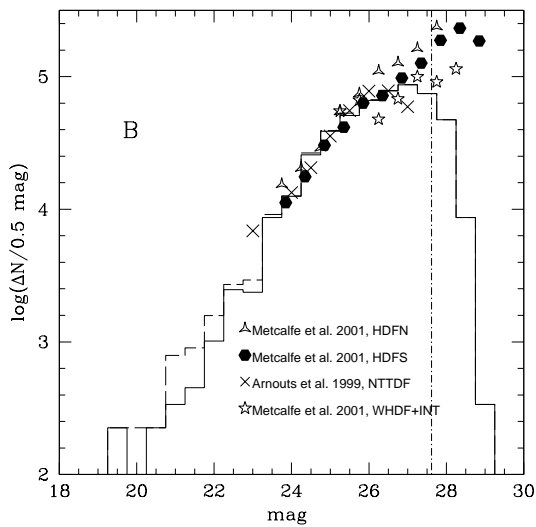,width=7.5cm,angle=0}
}}
\caption [B-band] {Galaxy number counts of the FDF in B band 
 (not corrected  for incompleteness) as compared
to other deep surveys. The vertical 
dash-dotted line indicates the 50\% completeness limits.}
\end{figure}

For the g-band, we just show our results in Fig. 7 without comparison,
since no adequate number counts have been presented in the literature 
for this passband.
Our estimated 50\% completeness limit is 26.86 mag in this filter.
\begin{figure}
\centerline{\hbox{
\psfig{figure=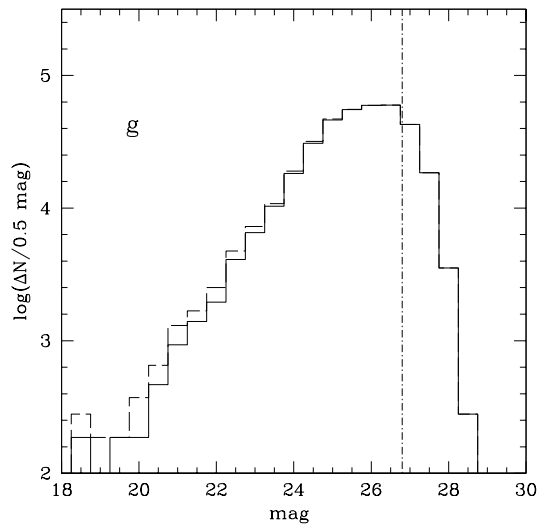,width=7.5cm,angle=0}
}}
\caption [g-band] {Galaxy number counts of the FDF in g band
(not corrected for incompleteness).
The vertical dash-dotted line indicates the 50\% completeness 
limits.}
\end{figure}

Our R-band and I-band data are  $50\%$-complete at $26.68$ mag and 
$26.37$ mag, respectively. Amplitude and slope agree well with
previously published fields. For the transformation of the HDF-counts
we followed Metcalfe et al. (\cite{whtdf})) and used $R\approx
R_{606,Vega}-0.1$ and $I\approx
I_{814,Vega}$; we also assumed that $R \approx R_{WHDF}$.
The counts are shown in Fig. 8 and 9.

\begin{figure}
\centerline{\hbox{
\psfig{figure=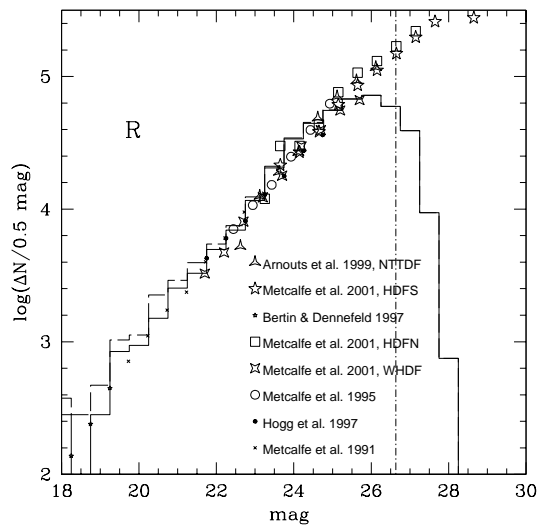,width=7.5cm,angle=0}
}}
\caption [R-band] {Galaxy number counts of the FDF in R band 
(not corrected for incompleteness) as compared
to other deep surveys. The vertical 
dash-dotted line indicates the 50\% completeness limits.}
\end{figure}
\begin{figure}
\centerline{\hbox{
\psfig{figure=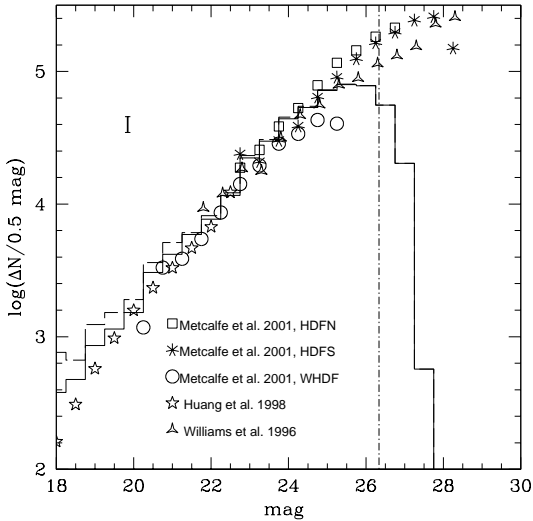,width=7.5cm,angle=0}
}}
\caption [I-band] {Galaxy number counts of the FDF in I band 
(not corrected for incompleteness)
as compared
to other deep surveys. The vertical 
dash-dotted line indicates the 50\% completeness limits.}
\end{figure}

Our number counts in the J-band (Fig. 10) agree  with those
derived by Saracco et al. (\cite{saracco}), and precisely match those of 
Teplitz et al. (\cite{teplitz}). The completeness is $22.85$ mag
and $23.60$ mag for the shallower and deeply 
exposed (factor of four in integration time) part of the field, respectively. 
Our number counts in the K-band  (Fig. 11) 
agree well with those of K\"ummel \& Wagner (\cite{kuemmel}) and  
Huang et al. (\cite{huang}). The
completeness limits are $20.73$ mag and $21.57$ mag for the shallow and deep 
exposed part of the field. 
For fairly shallow J and K pointings ($J\la 22$ and 
$K\la 20$) the field-to-field variations are expected to be significant
for our field size, since the distribution of massive, old systems
dominating the NIR frames varies considerably on small scales.
This has been demonstrated e.g. in the different pointings of the 
MUNICS survey by Drory et al. (\cite{drory}). 
The agreement with other surveys is good and the quoted detection limit 
correspond to the 50\% completeness limit of our sample. 

\begin{figure}
\centerline{\hbox{
\psfig{figure=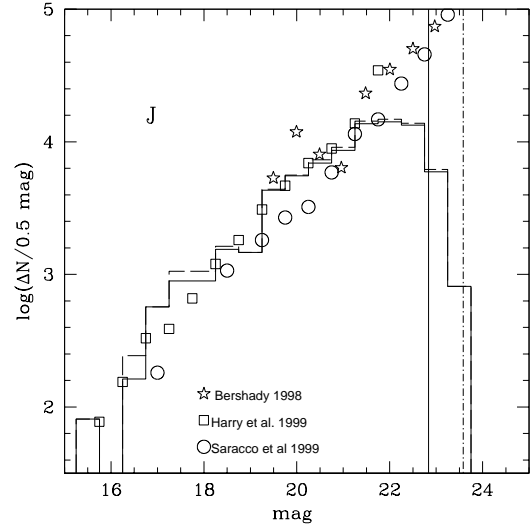,width=7.5cm,angle=0}
}}
\caption [J-band] {Galaxy number counts of the FDF in J band 
(not corrected for incompleteness) as compared
to other deep surveys. The vertical solid line indicates the 50\% completeness
for the shallower exposed part of the field, whereas the vertical 
dash-dotted line indicates the 50\% completeness for the deeply 
exposed part of the field.}
\end{figure}
\begin{figure}
\centerline{\hbox{
\psfig{figure=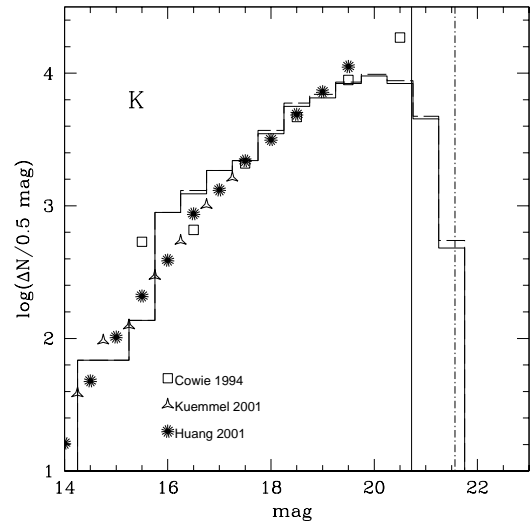,width=7.5cm,angle=0}
}}
\caption [K-band] {Galaxy number counts of the FDF in Ks band 
(not corrected for incompleteness) as compared
to other deep surveys. The vertical solid  line indicates the 50\% completeness
for the shallower exposed part of the field, whereas the vertical 
dash-dotted line indicates the 50\% completeness for the deeply 
exposed part of the field.
}
\end{figure}

\acknowledgements{We thank the Paranal and NTT staff at ESO for their
excellent and very efficient support at the telescope. 
We also thank the referee (Dr. M. Franx) for his constructive comments.
This work has been supported by the 
Deutsche Forschungsgemeinschaft (SFB 375, SFB 439), 
the VW foundation (I/76520) and the German Federal 
Ministry of Science and Technology
(Grants 05 2HD50A, 05 2GO20A and 05 2MU104).

We have made use of the Simbad Database, operated at CDS,
Strasbourg, France, and the NASA/IPAC Extragalactic Database (NED), operated
by the Jet Propulsion Laboratory, California institute of Technology under
contract with the National Aeronautics and Space Administration.}

\end{document}